\def\o{\over}
\def\A{\rightarrow}

\def\a{\alpha}
\def\b{\beta}

\def\k{\kappa}
\def\e{\epsilon}
\def\p{\pi}
\def\th{\theta}

\def\G{{\rm GeV}}

\def\E{{\rm eV}}
\documentstyle [12pt]{article}
\begin{document}
\baselineskip=25pt
\setcounter{page}{1}
\thispagestyle{empty}
\topskip 0.5  cm
\begin{flushright}
\begin{tabular}{c c}
& {\normalsize   EHU-1995-8}\\
& \today
\end{tabular}
\end{flushright}
\vspace{1 cm}
\centerline{\Large\bf Renormalization Effect on Large}
\centerline{\Large\bf  Neutrino Flavor Mixing in the}
\centerline{\Large\bf Minimal Supersymmetric Standard Model}
\vskip 1.5 cm
\centerline{{\bf Morimitsu TANIMOTO}
  \footnote{E-mail address: tanimoto@edserv.ed.ehime-u.ac.jp}}
\vskip 0.8 cm
 \centerline{ \it{Science Education Laboratory, Ehime University,
 790 Matsuyama, JAPAN}}
\vskip 1.5 cm
\centerline{\bf ABSTRACT}
\vskip 0.5 cm

  In the minimal supersymmetric standard model, we have studied the
 evolution of the neutrino flavor mixing
 by using the renormalization group equation(RGE)
 with  the Georgi-Jarlskog texture for the Yukawa coupling matrices.
For the large Yukawa coupling of the charged lepton, i.e.,
$\tan\b \gg 1$, the neutrino flavor mixing increases
significantly with running down to the electroweak scale by the RGE.
If one wishes to get the large neutrino flavor mixing $\sin \th_{23}$
at the electroweak scale, which is suggested by the  muon neutrino deficit
in the atomospheric neutrino flux,
 the initial condition $\sin \th_{23}\geq 0.27$
is required  at the GUT scale.
Combined with the see-saw enhancement of the neutrino flavor mixing,
  the large  mixing is naturally reproduced
 by setting  the reasonable initial condition.
\newpage
\topskip 0 cm
 The recent observed solar neutrino deficit[1] and   muon neutrino deficit
in the atomospheric neutrino flux[2] stimulate the systematic study of the
neutrino flavor mixings[3].
 In the standpoint of the quark-lepton unification in most
GUT groups, the Dirac mass matrix of neutrinos is similar to the one of
quarks, therefore, the neutrino flavor mixings turn out to be typically of
the same
order of magnitude as the quark mixings.
However, some authors pointed out that the
large neutrino flavor mixing  could be obtained in the see-saw mechanism[4]
as a consequence of certain structure of the right-handed
Majorana mass matrix[5,6,7]. That is the    so called see-saw
enhancement[7] of  the neutrino flavor mixing due to the cooperation
between the Dirac and
    Majorana mass matrices.
 The see-saw enhancement of the mixing
 gives a clue to solve the problem of the
  muon neutrino deficit in the atomospheric neutrino flux[2]
 since the deficit may be derived from  the large mixing
 between the  muon neutrino and other neutrino.
\par
   The see-saw enhancement condition[7] given by
Smirnov was modified in presence of the right-handed phases of the Dirac mass
matrix and the Majorana mass marix[8].
It was pointed out that the see-saw enhancement could be  obtained even if the
Majorana matrix is
proportional to the unit matrix, i.e., there is no hierarchy in the Majorana
mass matrix.  This enhancement is caused by the right-handed phases, which
never appear in the case of the quark mixing.
Thus, it seems that the large neutrino flavor mixing is naturally reproduced
 in  some GUT models with the see-saw mechanism.\par
There may be another enhancement mechanism of the neutrino flavor mixing.
Recently, the renormalization group equation(RGE) of the see-saw neutrino mass
operators with dimension-5 has been investigated by some authors[9,10].
Babu, Leung and Pantaleone
pointed out that the neutrino flavor mixing  is enhanced
 by the RGE in the minimal supersymmetric standard model(MSSM)[10].
  They have shown that  even resonant mixing occurs
 between the GUT scale and the electroweak scale
for  the large Yukawa coupling of the charged lepton, which corresponds to
  the case of $\tan\b\gg 1$ in the MSSM,
 where $\tan\b\equiv v_2/v_1$, $v_1$ and $v_2$ being of
 the vacuum expectation values with $v_1^2+v_2^2=v^2=(174\G)^2$.
 Thus, it is important to investigate the enhancement  of the neutrino
 flavor mixing by the RGE  in the case of  the large Yukawa couplings
of the charged lepton.
In this paper, we study how large  the neutrino  flavor mixing
 is enhanced by the RGE in the realistic model.
\par
We start with writting the RGE for the Yukawa couplings
 ${\bf Y}_U$, ${\bf Y}_D$, ${\bf Y}_E$ and ${\bf Y}_N$,
which are $3\times 3$ matrices
 in the generation space.  Since we are concerned with the case of the large
Yukawa couplings, we neglect the first generation.
 Therefore, those Yukawa coupling matrices shall be considered
 to be $2\times 2$ matrices.
We have the simple RGE at one-loop approximation in the MSSM as follows[11,12]:
 \begin{eqnarray}
  8\p^2 {d\o dt}({\bf Y}_U {\bf Y}_U^\dagger)& = &
 \{ -\sum_i c_U^i g_i^2+3 ({\bf Y}_U {\bf Y}_U^\dagger)
+{\rm Tr}[3({\bf Y}_U {\bf Y}_U^\dagger)+({\bf Y}_N {\bf Y}_N^\dagger)] \}
({\bf Y}_U {\bf Y}_U^\dagger)  \nonumber \\
       & &\qquad \qquad   +{1\o 2}\{({\bf Y}_D {\bf Y}_D^\dagger)
 ({\bf Y}_U {\bf Y}_U^\dagger)+({\bf Y}_U {\bf Y}_U^\dagger)
({\bf Y}_D {\bf Y}_D^\dagger)\}       \ , \nonumber \\
  8\p^2 {d\o dt}({\bf Y}_D {\bf Y}_D^\dagger)& = &
 \{ -\sum_i c_D^i g_i^2+3 ({\bf Y}_D {\bf Y}_D^\dagger)
 +{\rm Tr}[3({\bf Y}_D {\bf Y}_D^\dagger)+({\bf Y}_E {\bf Y}_E^\dagger)] \}
({\bf Y}_D {\bf Y}_D^\dagger)  \nonumber \\
       & &\qquad \qquad   +{1\o 2}\{({\bf Y}_U {\bf Y}_U^\dagger)
 ({\bf Y}_D {\bf Y}_D^\dagger)+({\bf Y}_D {\bf Y}_D^\dagger)
({\bf Y}_U {\bf Y}_U^\dagger)\}       \ , \nonumber \\
  8\p^2 {d\o dt}({\bf Y}_N {\bf Y}_N^\dagger)& = &
 \{ -\sum_i c_N^i g_i^2+3 ({\bf Y}_N {\bf Y}_N^\dagger)
+{\rm Tr}[3({\bf Y}_U {\bf Y}_U^\dagger)+({\bf Y}_N {\bf Y}_N^\dagger)] \}
({\bf Y}_N {\bf Y}_N^\dagger)  \nonumber \\
       & &\qquad \qquad   +{1\o 2}\{({\bf Y}_E {\bf Y}_E^\dagger)
 ({\bf Y}_N {\bf Y}_N^\dagger)+({\bf Y}_N {\bf Y}_N^\dagger)
({\bf Y}_E {\bf Y}_E^\dagger)\}       \ , \nonumber \\
  8\p^2 {d\o dt}({\bf Y}_E {\bf Y}_E^\dagger)& = &
 \{-\sum_i c_E^i g_i^2+3 ({\bf Y}_E {\bf Y}_E^\dagger)
+{\rm Tr}[3({\bf Y}_D {\bf Y}_D^\dagger)+({\bf Y}_E {\bf Y}_E^\dagger)] \}
({\bf Y}_E {\bf Y}_E^\dagger)  \nonumber \\
       & &\qquad \qquad   +{1\o 2}\{({\bf Y}_N {\bf Y}_N^\dagger)
 ({\bf Y}_E {\bf Y}_E^\dagger)+({\bf Y}_E {\bf Y}_E^\dagger)
({\bf Y}_N {\bf Y}_N^\dagger)\}       \ ,
 \end{eqnarray}
\noindent with
\begin{equation}
 t=\ln\left ( \mu \right ) \ ,
\end{equation}
\noindent
where the coefficients  $c^i_U$,  $c^i_D$,
$c^i_N$ and $c^i_E(i=1,2,3)$ are given to be
 (13/15, 3, 16/3), (7/15, 3, 16/3), (3/5, 3, 0) and (9/5, 3, 0),
respectively  in the MSSM.
\par
However, the neutrino Yukawa coupling ${\bf Y}_N$ decouples from
 the RGE in eq.(1)  below the right-handed Majorana mass scale, in which
the heavy right-handed neutrinos decouple,
 and then, the evolution equation of the neutrino Yukawa coupling ${\bf Y}_N$
 becomes meaningless.
Recently, the RGE of the see-saw neutrino mass operator
 with the dimension-5
have been investigated by some authors.
 Below the right-handed Majorana mass scale, we use the RGE of the
 neutrino mass operator given by Babu, Leung and Pantaleone[10]
in the MSSM, which is written   as follows:
 \begin{equation}
    8\p^2 {d\o dt}{\bf \k}_N = \{-({3\o 5} g_1^2+3 g_2^2)+
{\rm Tr} [3{\bf Y}_U {\bf Y}_U^\dagger]\}\k_N
 +{1\o 2}\{({\bf Y}_E {\bf Y}_E^\dagger)\k_N +
   \k_N ({\bf Y}_E {\bf Y}_E^\dagger)^T\}       \ ,
 \end{equation}
\noindent where
  \begin{equation}
      {\bf \k}_N \simeq {\bf Y}_N {\bf M}_R^{-1}{\bf Y}_N^T  \ ,
 \end{equation}
\noindent
where ${\bf M}_R$ is a right-handed Majorana matrix, which is
for a while assumed to be proportional to the unit matrix
 such as ${\bf M}_R=M_R {\bf I}$.  Hereafter,  $M_R$  refers to the
scale of the right-handed Majorana Mass.
Then, the neutrino mixing angle $\th_{23}$
 for the third generation and the second one
is determined by the evolution equation,
 \begin{equation}
    16\p^2 {d\o dt}{\sin^2 2\th_{23}} =-2\sin^2 2\th_{23}
        (1-\sin^2 2\th_{23})
 (Y_{E3}^2-Y_{E2}^2){(\k_{N})_{33}+(\k_{N})_{22}\o
(\k_{N})_{33}-(\k_{N})_{22}}  \ ,
 \end{equation}
\noindent
  where $Y_{E3}$ and $Y_{E2}$ are the Yukawa couplings of
the third generation charged lepton and the second generation one,
 respectively.
It is noticed that $\sin^2 2\th_{23}$ increases with running down
from the GUT scale ($M_{GUT}$)  to the electroweak scale ($M_Z$).
 In particular, the evolution of the
mixing angle is remarkable if $Y_{E3}$ is large.
For example,  if
 $Y_{E3}=3$ is fixed between the $M_{GUT}$ scale and the $M_Z$
 one, $\sin^2 2\th_{23}$ can reach the maximal value $1$
 on the way to the $M_Z$ scale[10].
\par
 However, the magnitude of the Yukawa coupling of the charged lepton
  rapidly decreases  in the evolution of the RGE as seen in eq.(1) even if
 the  large Yukawa coupling is put  at the $M_{GUT}$ scale.
We show the evolution of  $Y_{E3}$ for
 $Y_{E3}=3,  2,  1, 0.5$ at the $M_{GUT}$ scale in fig.1,
 where we have neglected Yukawa couplings of the second generation and
 we have taken  GUT conditions $Y_{D3}=Y_{E3}$ and $Y_{U3}=Y_{N3}=3$
 with $M_{GUT}=3\times 10^{16}\G$, $M_{R}=3\times 10^{14}\G$
and $\a_s(M_Z)$=0.125[13].
The dashed-lines denote evolutions from $M_{GUT}$ to
 $M_{R}$
 and the solid lines denote evolutions from $M_{R}$ to
  $M_Z$.
 Thus, it is not clear whether the neutrino flavor mixing is enough enhanced.
 Therefore,  we study quantitatively the enhancement of $\sin\th_{23}$
due to the large Yukawa coupling of the charged lepton
   in the specified model of the Yukawa coupling matrices.  \par
\begin{center}
\unitlength=0.7 cm
\begin{picture}(2.5,2.5)
\thicklines
\put(0,0){\framebox(3,1){\bf fig. 1}}
\end{picture}
\end{center}

In our studies, we take the Georgi-Jarlskog texture[14] for
the Yukawa coupling matrices at the $M_{GUT}$ scale  since this texture
 is intensively investigated in the quark- and the lepton-mixings[3,15].
It was concluded that the acceptable solution of the quark mixings
 lie at the edges of allowed regions in this texture.
The Yukawa coupling matrices are written in the three generation space as
 follows:
 \begin{equation}
  {\bf Y}_U =  \left( \matrix{0 & C_U  & 0\cr C_U  & 0 &  B_U \cr
       0 & B_U  & A_U \cr} \right )     , \ \
  {\bf Y}_D =  \left( \matrix{0 & C_D  & 0\cr C_D  & B_D &  0 \cr
       0 & 0  & A_D \cr} \right )     , \ \
  {\bf Y}_E =  \left( \matrix{0 & C_D  & 0\cr C_D  & -3 B_D &  0 \cr
       0 & 0  & A_D \cr} \right )     , \ \
  \end{equation}
\noindent for the up-quarks, down quarks and the charged leptons.
In this texture with $SU(5)$,
 the neutrino Yukawa couplings are not related to
 the up-quark ones. In SO(10), the neutrino Yukawa couplings could be related
 to the up-quark ones. Then, the neutrino Yukawa coupling matrix
 has same form as  the up-quark one in eq.(6).  However,
 the magnitudes of the matrix elements depend on
 the multiplets of the Higgs fields such as $\bf \underline {10}$ or
 $\bf \underline {126}$. If the vacuum expectation values of
 both multiplets contribute to the  matrix elements,
  the magnitudes of the matrix elements
 are independent each other in the up-quark sector and the neutrino sector.
 In general, the matrix of the Yukawa coupling is written as,
 \begin{equation}
  {\bf Y}_N =  \left( \matrix{0 & C_N  & 0\cr C_N  & 0 &  B_N \cr
       0 & B_N  & A_N \cr} \right )    \ .
 \end{equation}
\noindent
Hereafter, we consider only the third generation and the second one
 in the matrices of eqs.(6) and (7)
 since our concern is the mixing angle in the
case of the large Yukawa couplings.
Then, the free parameters  are matrix elements
 $A_U$, $B_U$, $A_N$, $B_N$ $A_D$ and $B_D$, and the
 right-handed Majorana mass  $M_R$.
\par
At first, the parameter $A_D$ is taken to be as possible as
 large in order to study how large
  the neutrino flavor mixing  is enhanced by the evolution of
 the RGE in the case of
 the large Yukawa coupling of the charged lepton.
 In the framework of the pertubative unification, we take a possible
large value $A_D$, while the large top quark mass
 favours the large $A_U$.  Then, $A_N$ is also
 expected to be large due to the quark-lepton symmetry.
For definiteness, we take the following values in our calculation:
 \begin{equation}
  A_U=A_N=A_D=3 .
 \end{equation}

 The unification of
 $b$-quark and $\tau$-lepton masses at the $M_{GUT}$ scale,
 where the three gauge couplings unify,
 is one of crucial issues of the Grand Unification.
 In the MSSM, there is a successful gauge coupling
unification[16] and also the $b-\tau$ unification is possible.
Vissani and Smirnov[12] studied the $m_b/m_\tau$ ratio
in the standpoint of the $b-\tau$ unification in the MSSM.
 They showed that the ratio
 depends crucially on  $\a_s(M_Z)$ and $M_R$ as well as the Yukawa couplings.
The increase of $A_U$ or $A_D$ tends to decrease the ratio.
The gauge coupling $\a_s(M_Z)$ works in opposite direction.
The neutrino renormalization also results in increase of the ratio.
In other words, the decrease of $M_R$ and the increase of $A_N$
 tend to increase the ratio.
 Our input values in eq.(8)  predict
 the rather small  $m_b/m_\tau$ ratio.
Fixing $M_{GUT}=3\times 10^{16}\G$, $M_{R}\simeq 3\times 10^{14}\G$
and $\a_s(M_Z)=0.125$, our predicted ratio is $m_b/m_\tau=1.54$
at the $M_{Z}$ scale. This value corresponds
 to the pole mass of the $b$-quark $m^{\rm pole}_b\simeq 4.75\G$,
 which is derived by using  the two-loop standard model RGE
 with  $\a_s(M_Z)=0.125$ and $m_\tau=1.7771\G$[17].
  This  value of the pole mass
  is consistent  with the experimental value.
 Actually,
Dominguez and Paver have given $m^{\rm pole}_b\simeq 4.72\pm 0.05\G$[18]
 although  Titard and  Yndur\'ain have obtained
 the rather large value $4.9\G$[19].
  It may be  useful for readers to add a following comment:
    one could get  $m^{\rm pole}_b=4.85\G$  by changing the parameters such as
  $A_N=3\A 5$ or  $\a_s(M_Z)=0.125\A 0.127$.
 \par
 Now, the flavor mixings of the quark sector and
the lepton sector can be investigated by the use of
 the parameters $B_U$ and $B_N$.
By taking $B_U=0.086$, which gives $\sin \th_{23}=0.029$ at $M_{GUT}$,
 we get the quark mixing,
$\sin \th_{23}(M_Z)=0.051$, which may lie just at the edge of the
 experimentally allowed region of $V_{cb}$[20] as discussed in
 ref.15.
On the other hand, $B_N$ is a free parameter for the present.
 In order to investigate the evolution of the neutrino flavor mixing,
 we take $B_N=0.65,\ 0.83,\ 1.05$, which  correspond to
 the neutrino flavor mixing
 $\sin \th_{23}(M_{GUT})=0.2,\  0.25,\ 0.3$, respectively.
We show the evolution of the neutrino flavor mixing
 $\sin^2 2\th_{23}$ in fig.2, in which
 $M_{GUT}=3\times 10^{16}\G$, $M_{R}= 2.2\times 10^{14}\G$
and $\a_s(M_Z)=0.125$ are taken.
Then, we get $m_t(M_Z)=177\G$ and $\tan\b=60.8$,
which are given by
 \begin{equation}
   m_t(M_Z)=Y_t(M_Z)  {\tan\b \o \sqrt{1+\tan^2\b}} v \ , \qquad
   m_{\tau}(M_Z)=Y_{\tau}(M_Z)  {1\o \sqrt{1+\tan^2\b}} v \ .
 \end{equation}
\noindent
The predicted $m_t(M_Z)$ corresponds to  $m_t^{\rm pole}=188\G$,
 which is consistent with the recent experimental measurements[21].
The  increases of  $\sin^2 2\th_{23}$ are
 $0.08$($53\%$),  $0.11$($48\%$) and $0.14$($43\%$) at the $M_Z$ scale
for each intial condition $\sin^2 2\th_{23}=0.16$, $0.24$ and $0.33$,
respectively, at the $M_{GUT}$ scale.
The evolution is remarkably rapid near the  $M_{GUT}$ scale,
 but becomes  rather mild below the $M_R$ scale.
This behaviour is understandable if one finds out the behaviour of the
 evolution of the Yukawa coupling of the charged lepton in fig.1.
Thus, it is impossible to get the remarkable enhancement
 of the neutrino mixing to reproduce $\th_{23}(M_Z)=\pi/4$
 unless we set the large neutrino flavor mixing
as an initial condition at $M_{GUT}$ in the MSSM.
\begin{center}
\unitlength=0.7 cm
\begin{picture}(2.5,2.5)
\thicklines
\put(0,0){\framebox(3,1){\bf fig. 2}}
\end{picture}
\end{center}

However, our result may present  a clue to solve the problem of
the muon neutrino deficit in the atomospheric neutrino flux.
The atomospheric neutrino deficit data  presented by KAMIOKANDE and IBM[2]
 have given under the assumption of the dominant mixing
 between the muon neutrino and the tau neutrino as follows:
\begin{equation}
 \Delta m_{23}^2=(0.3\sim 3)\times 10^{-2} {\rm eV}^2  \ ,
\qquad \sin^2 2\th_{23}=0.4\sim 1 \ ,
 \end{equation}
\noindent
where  rather conservative bounds are taken.
In fig.2, the predicted mass of the tau neutrino is fixed such as
 $m_{\nu_{\tau}}=0.1 {\rm eV}$ by taking $M_{R}=2.2\times 10^{14}\G$.
In order to get $\sin^2 2\th_{23}\geq 0.4$ at $M_Z$, one should prepare the
initial condition  $\sin^2 2\th_{23}(M_{GUT})\geq  0.27$.\par
Here, we comment on our prediction of the
 muon neutrino mass $m_{\nu_{\mu}}$.  The predicted mass
  is  $3.2\times 10^{-3} {\rm eV}$ in the case of
$\sin^2 2\th_{23}\simeq 0.4$ at the $M_Z$ scale. This predicted value is
consistent with
 the MSW solution of the solar neutrino data(small mixing angle solution)
[1,22],
 \begin{equation}
 \Delta m^2_{12}=(0.5\sim 1.2)\times  10^{-5}{\rm eV}^2 \ .
 \end{equation}
\noindent
The mixing angle between the muon neutrino and the electron neutrino
 is beyond the scope of our model because the first generation
 is neglected.\par
 In above calculations, the parameter $B_N$ is a free parameter.
However, this parameter may be related with the one in the quark sector.
Here, we study the specific cases of
 $B_N=B_U=0.086$ and $B_N=0.64$. The former case denotes
the quark-lepton unification of the mass matrix elements, and the latter one
 means   $\sin \th_{23}=0.2$, i.e.,  around the Cabibbo angle
 at the $M_{GUT}$ scale.
Of course, these values of $B_N$ are not enough large
 to get $\sin^2 2\th_{23}\geq 0.4$ at the $M_Z$ scale.
Until now, we have  assumed that the right-handed Majorana
matrix ${\bf M}_R$ is proportinal to the unit matrix.
However, this assumption may be modified simply
in the two generation space as follows:
 \begin{equation}
{\bf M}_R = \left( \matrix{ M_{R2} &  0 \cr  0  & M_{R3} \cr} \right )
 = M_R \left( \matrix{ \e_R &  0 \cr  0  & 1 \cr} \right ) \ ,
\end{equation}
\noindent where $M_R\equiv M_{R3}$ and $\e_R\equiv M_{R2}/ M_{R3}$.
The parameter $\e_R$ is a complex number due to the
 unknown right-handed phase in general.
Of course, this matrix could have off-diagonal elements,
but we do not discuss the case.
The diagonal matrix in eq.(12) is enough to investigate the see-saw
enhancement. Then, at the GUT scale the light neutrino flavor
mixing angle is given  as follows:
 \begin{equation}
\tan{2 \th_{23}} = {2A_N B_N \o A_N^2-B_N^2+{B_N^2\o \e_R}} \ \  .
\end{equation}
\noindent The see-saw enhancement is obtained if $\e_R\simeq -B_N^2/A_N^2$
 is taken[7,8].
 Now, we study the RGE evolution  of the neutrino flavor mixing
 with  the see-saw  enhancement. \par
 Let us begin with discussing the case  of $B_N=B_U=0.086$.
 If we take $\e_R=1$, we get the very small  initial value
$\sin \th_{23}=0.029$ at the $M_{GUT}$  scale.
It is impossible to obtain $\sin^2 2\th_{23}\geq 0.4$ at the $M_Z$ scale
 by using this initial condition.
 Therefore, we need  the significant see-saw enhancement
  in the initial condition.
 We show the evolution with $\e_R=-1/800,\  -1/950,\ -1/1050$ in fig.3(a),
 in which $M_R=3\times 10^{13}\G$ is taken.
 The initial conditions are
 $\sin^2 2\th_{23}(M_{GUT})=0.027, 0.064, 0.150$
 for $\e_R=-1/800,\  -1/950,\ -1/1050$, respectively.
 If we take $\e_R=-1/1050$, we get $\sin^2 2\th_{23}(M_Z)=0.48$, which is
 much enhanced from the initial value.
The neutrino masses are
 $m_{3}=9.6\times 10^{-2}\E$ and $m_{2}=1.5\times 10^{-2}\E$,
 in which $m_3$ is consistent with the atmospheric neutrino data in eq.(10),
 however, $m_2$ is too large to explain the solar neutrino data
 in eq.(11).
 This enhancement of the neutrino flavor mixing
is caused by the see-saw mechanism at the initial condition
 and then, the enhanced mixing increses by the RGE.
\par
\begin{center}
\unitlength=0.7 cm
\begin{picture}(2.5,2.5)
\thicklines
\put(-2,0){\framebox(6,1){\bf fig. 3(a) and 3(b)}}
\end{picture}
\end{center}

The large mass ratio $|M_{R3}/M_{R2}|\simeq 1000$ may not be
reasonable in  a realistic model.
In the case of $B_N=0.64$, we do not need such large mass ratio
 in the Majorana mass matrix.
We show the evolution with $\e_R=-1/2,\  -1/4,\ -1/5$ in fig.3(b),
 in which $M_R=3.2\times 10^{14}\G$ is taken.
 The enhanced initial conditions are
 $\sin^2 2\th_{23}(M_{GUT})=0.196, 0.234, 0.256$
 for $\e_R=-1/2,\  -1/4,\ -1/5$, respectively.
If we take $\e_R=-1/5$,  $\sin^2 2\th_{23}$ increses to $0.47$
 at the $M_Z$ scale.
The neutrino masses are
 $m_{3}=5.5\times 10^{-2}\E$ and $m_{2}=3.5\times 10^{-3}\E$,
 which lie at the edges of allowed regions eqs.(10) and (11).
  If we get $\sin^2 2\th_{23}(M_{Z})\geq 0.47$ by taking
 $|\e_R|\leq 1/5$,
  the muon neutrino mass $m_2$ becomes comparable to the tau neutrino mass
 $m_3$. Then, the solar neutrino data cannot be explained.
Since the mass ratio $|M_{R3}/M_{R2}|= 2\sim 5$ is  a reasonable one
 in the right-handed Majorana mass matrix, this case may be a realistic one.
\par
The summary is given as follows.
In the MSSM, we have studied the evolution
 of the neutrino flavor mixing by using the RGE with  the Georgi-Jarlskog
texture for the Yukawa coupling matrices.
For the large Yukawa coupling of the charged lepton, i.e.,
$\tan\b \gg 1$, the neutrino flavor mixing increases
significantly with running down to the electroweak scale
from the GUT scale by the RGE.
However, the maximal flavor mixing could not be
 reproduced unless the large mixing is  set as an
 intial condition at the GUT scale.
If one wishes to get the large neutrino flavor mixing at the $M_Z$ scale,
 which is suggested by the  muon neutrino deficit
in the atomospheric neutrino flux,
 the initial condition $\sin \th_{23}(M_{GUT})\geq 0.27$
is required  at the GUT scale.
Combined with the see-saw enhancement of the neutrino flavor mixing,
  the large  mixing is reproduced by the evolution of the RGE
 using the input of  the realistic parameters such as
$A_N/B_N\leq 3/0.64$ at the GUT scale
 and $|M_{R3}/M_{R2}|\geq 5$.
Although our analyses have been done by using  the Georgi-Jarlskog
texture for the Yukawa coupling matrices,
 our qualitative conclusion does not change even if other texture will be
  used.
The RGE evolution of the neutrino flavor
mixing should be taken into consideration
 seriously in building  the  model with the large neutrino flavor mixing.
\par
\vskip 1 cm
\centerline{\bf Acknowledgments}
This research is supported by the Grant-in-Aid for Science Research,
Ministry of Education, Science and Culture, Japan(No. 07640413).
\newpage
\centerline{\large \bf References}
\vskip 1 cm
\noindent
[1]  GALLEX Collaboration,  Phys. Lett.  {\bf 327B}(1994)377;\par
 SAGE Collaboration, Phys. Lett.  {\bf 328B}(1994)234;\par
 Homestake Collaboration, Nucl. Phys. {\bf B38}(Proc. Suppl.)(1995)47;\par
Kamiokande Collaboration, Nucl. Phys. {\bf B38}(Proc. Suppl.)(1995)55.
\par\noindent
[2] K.S. Hirata et al., Phys. Lett. {\bf 280B}(1992)146;\par
 R. Becker-Szendy et al,. Phys. Rev.
      {\bf D46}(1992)3720;\par
 Y. Fukuda et al., Phys. Lett. {\bf 335B}(1994)237.\par\noindent
[3] C.H. Albright, Phys. Rev.
     {\bf D43}(1991)R3595, {\bf D45}(1992)R725;\par
 G. Lazaridis and Q. Shafi, Nucl. Phys. {\bf B350}(1991)179;\par
 S.A. Bludman, D.C. Kennedy and P. Langacker, Nucl. Phys.
{\bf B373}(1992)498;\par
 K.S. Babu and Q. Shafi, Phys. Lett. {\bf 294B}(1992)235;\par
 S. Dimopoulos, L. Hall and S. Raby, Phys. Rev. Lett.
   {\bf 68}(1992)1984; \par
  Phys. Rev. {\bf D45}(1992)4192;\par
 M. Fukugita, M. Tanimoto and T. Yanagida, Prog. Theor. Phys.
     {\bf 89}(1993)263;\par
M. Tanimoto, Mod. Phys. Lett. {\bf 9A}(1994)1827.\par
\noindent
[4] M. Gell-Mann, P. Ramond and R. Slansky, in {\it Supergravity},
 Proceedings of the \par
Workshop, Stony Brook, New York, 1979, edited by P. van Nieuwenhuizen \par
and D. Freedmann, North-Holland, Amsterdam, 1979, p.315;\par
 T. Yanagida, in {\it Proceedings of the Workshop on the Unified Theories and
Baryon\par
 Number in the Universe}, Tsukuba, Japan, 1979, edited by O. Sawada and \par
A. Sugamoto, KEK Report No. 79-18,
    Tsukuba, 1979, p.95.\par\noindent
[5] M. Tanimoto, T. Hayashi and M. Matsuda, Zeit. f\"ur Physik,
   {\bf C58}(1993)267.\par\noindent
[6] C.H. Albright and S. Nandi,  Fermilab-Pub-94/061-T
      (1994)(unpublished).\par\noindent
[7] A. Yu. Smirnov, Phys. Rev. {\bf D48}(1993)3264;
      IC/93/359(1993)(unpublished).\par\noindent
[8] M. Tanimoto, Phys. Lett. {\bf 345B}(1995)477.\par
\noindent
[9] P.H. Chankowski and Z. Pluciennik, Phys. Lett.
    {\bf 316B}(1993)312.\par
\noindent
[10] K.S. Babu, C.N. Leung and J. Pantaleone,
   Phys. Lett.  {\bf 319B}(1993)191.\par
  \noindent
[11] N.K. Falck, Zeit. f\"ur Physik, {\bf C30}(1986)247.\par
\noindent
[12] F. Vissani and A.Y. Smirnov, Phys. Lett. {\bf 341B}(1994)173.\par
\noindent
[13] In order to get the reasonable $b$-quark pole mass,
  we prefer  $\a_s(M_Z)=0.125$,\par
 which was   given by the LEP data,
to  the world average value \par
 $\a_s(M_Z)=0.117\pm 0.005$  in Ref.20.
\par\noindent
[14] H. Georgi and C. Jarlskog, Phys. Lett. {\bf 86B}(1979)297.\par
\noindent
[15] V. Barger, M.S. Berger, T. Han and M. Zralek, Phy. Rev. Lett.
        {\bf 68}(1992)3394.\par
\noindent
[16] U. Amaldi, W. de Boer and H. F\"urstenau,
        Phys. Lett. {\bf 260B}(1991)447;\par
 P. Langacker and N. Polonsky, Phys. Rev. {\bf D47}(1993)4028.\par
\noindent
[17] H. Arason, D.J. Casta\~no, B. Kesthelyi, S. Mikaelian, E.J. Piard,
   P. Ramond \par
  and   B.D. Wright, Phys. Rev. {\bf D46}(1992)3945.
 \par\noindent
[18] C.A. Dominguez and N. Paver, Phys. Lett. {\bf 293B}(1992)197.\par
\noindent
[19] S. Titard and F.J. Yndur\'ain, Phys. Rev. {\bf D49}(1994)6007.
\par
\newpage
\noindent
[20] Particle Data Group, Phys. Rev. {\bf  D50}(1994)1315.
     \par
\noindent
[21] CDF Collaboration, F. Abe et. al., Phys. Rev. Lett.
     {\bf  74}(1995)2626;\par
     D0 Collaboration, S. Abachi et.al., Phys. Rev. Lett.
     {\bf  74}(1995)2632.\par
\noindent
[22] S.P. Mikheyev and A.Yu. Smirnov, Yad. Fiz.
     {\bf 42}(1985)1441;\par
  L. Wolfenstein, Phys. Rev. {\bf  D17}(1987)2369;\par
 E.W. Kolb, M.S. Turner and T.P. Walker,
   Phys. Lett. {\bf 175B}(1986)478;\par
 S.P. Rosen and J.M. Gelb, Phys. Rev.
         {\bf D34}(1986)969;\par
   J.N. Bahcall and H.A. Bethe, Phys. Rev. Lett.
     {\bf  65}(1990)2233;\par
N. Hata and P. Langacker, Phys. Rev. {\bf  D50}(1994)632;\par
 P.I. Krastev and A. Yu. Smirnov, Phys. Lett. {\bf 338B}(1994)282.
\par
\newpage
\topskip 1 cm
\centerline{\large{\bf Figure Captions}}
\vskip 1.5 cm
\noindent
{\bf Fig.1}:\par
Running of the Yukawa coupling of the charged lepton.
The intial values are
 set to be $3$, $2$, $1$ and $0.5$, respectively.
The dashed-lines denote evolutions from $M_{GUT}$ to
 $M_{R}$  and the solid lines denote evolutions from $M_{R}$ to $M_Z$.
\par
\vskip 0.7 cm
\noindent
{\bf Fig.2}:\par
Running of the neutrino flavor mixing.
The intial values are set to be
$\sin \th_{23}=0.3$, $0.25$, and $0.2$, respectively at the $M_{GUT}$ scale.
The notations are same as in fig.1.\par
\vskip 0.7 cm
\noindent
{\bf Fig.3}:\par
Running of the neutrino flavor mixing.
The  parameter $\e_R$ is taken to be
\par \noindent
(a) $-1/1050$, $-1/950$ and $-1/800$,  and
(b) $-1/5$, $-1/4$ and $-1/2$, respectively.
\par \noindent
The notations are same as in fig.1.\par

\end{document}